\documentclass[aps,twocolumn,groupedaddress,showpacs,%
eqsecnum,secnumarabic,amssymb]{revtex4}

\usepackage[dvips]{graphicx}
\usepackage{bm}
\usepackage{amsmath}

\begin{document}

\title{The calculation of parameters for Stevens Hamiltonian
 in a crystalline field generated by the electric charge
 uniformly extended in one direction}
\author{V. Bobrovskii}
\email[Electronic address:]{bobrovskii@imp.uran.ru}
\author{A. Mirmelstein}
\author{A. Podlesnyak}
\author{I. Zhdakhin}
\affiliation{Institute for Metal Physics,
Ural Division of the Russian Academy of Sciences \\
620219 Ekaterinburg, Russia }
\date{\today}

\begin{abstract}
Exact expressions for the parameters of Stevens Hamiltonian are
derived within the framework of a specific model that assumes
uniform character of charge density distribution in a certain
direction over crystalline lattice.
\end{abstract}
\pacs{}
\maketitle

\section{Formulation of the model}

The well-known point charge model is widely used both
for description of crystal electric field (CEF) effects
and for analysis of the related experimental results.
A standard form of its application is based
on the so-called equivalent Stevens Hamiltonian
(ESH)~\cite{Hutchings,Bleaney,Wallace,Lea}.
The respective coefficients for the Hamiltonian have been
tabulated already for many crystal structures. It should be
noted that this Hamiltonian is not directly relevant
to the point character of the electric charges forming CEF.
Therefore, methods to calculate the Hamiltonian coefficients
can be generalized for a more complicated case.
For instance, in~\cite{Bobrovskii} they are determined
with taking into
account of screening effects by using of ligand electrostatic
potential in the form of Yukawa.
The numerical methods for calculation of the ESH parameters
are also in use. In particular, it is possible to advance
point charge model for more realistic assuming about
electron density distribution by introducing
a net of negatively charged points in interion
space~\cite{Podlesnyak}.
(This approach will be described in details in item 6 of the
present paper.)

The perovskite-related compounds comprising
CuO$_2$-planes in the structure were a
subject of intence studies during the last decade.
It is of a general believement that the charge state of these
planes is responsible for the unusual properties of the
layered copper perovskites as, for instance, the
high-temperature superconductivity is.
The inelastic neutron scattering has proved to be a powerful
tool for studying the charge states in the
high-temperature superconductors containing rare-earth
elements with f-shells~\cite{Mirmelstein88,Mesot,Mirmelstein97}.
A numerical analysis of the data collected in
the scattering experiments enabled us to obtain earlier some
evidences in proof of extended,
stripe-like charge structures in the
CuO$_2$-planes~\cite{Mirmelstein00}.
When analyzing these data we have found rather general
method to obtain exact solutions for ESH
parameters of a crystal field generated by the charge system
with density being uniform in one direction.

The specific goal of this paper is the systematic description
of the mathematical apparatus developed for analytical
calculation of ESH parameters for this geometry of charge
distribution and also the analysis of the consequences resulted
from the model.

Since the charge systems described by our theory are assumed
to have sufficiently high symmetry (at least orthorhombic)
it is convenient to use the local coordinate
system $(\tilde x, \tilde y, \tilde z)$
shown at Fig.~\ref{fig:1} with the origin at the rare
earth ion and the electric charges situated within two
parallel planes $\tilde z = \pm H$.
The density of the uniformly extended in $\tilde x$ direction
charge can be described by the function $f(L)$
where $L$ is the distance in $\tilde y$-direction between
$\tilde z$-axis and the partial charge
filament \footnote{Traditially notations $X,Y,Z$ are
used for coordinates of the ligands but
we would like to reserve $Y$ and
$Z$ for spherical and tesseral harmonics. To avoid a mishmach
we'll use for coordinates of the crystal electric field sources
in $(\tilde x, \tilde y, \tilde z)$-system the capital letters
$D,L,H$ respectively.}.

At first let us consider the electrostatic potential generated
by the infinite uniformly charged filament oriented
along $\tilde x$-axis. If the rare-earth ion has
f-electron at $\tilde x, \tilde y, \tilde z$
then the perturbing crystalline potential
generated by the filament will be:
\begin{equation}
U(\tilde x, \tilde y, \tilde z)= U(\tilde y, \tilde z) =
\xi \ln \frac{H^2 + L^2}{(L - \tilde y)^2 + (H - \tilde z)^2};
\label{eq:1.1}
\end{equation}
where $\xi$ is the linear charge density.

Taking into account that due to the small radii of the f-shell
the inequalities $|\tilde y|;|\tilde z| \ll \sqrt{L^2 + H^2}$
are satisfied and introducing the notations:
\begin{eqnarray}
\alpha & = & \frac{\tilde y}{\sqrt{L^2 + H^2}}; \quad
\beta =  \frac{\tilde z}{\sqrt{L^2 + H^2}}; \nonumber \\
M & = & - \frac{2L}{\sqrt{L^2 + H^2}}; \quad
N = - \frac{2H}{\sqrt{L^2 + H^2}};
\label{eq:1.2}
\end{eqnarray}
one can present the expression (\ref{eq:1.1}) as a polynomial
expansion:
\begin{equation}
U(\tilde y, \tilde z)  = \xi \sum_{p,q} F_{p,q} \alpha^p \beta^q.
\label{eq:1.3}
\end{equation}
As a consequence of the selection rules for the ESH matrix
elements \cite{Hutchings}, the non-zero contribution into
ESH is produced only by the members of expansion
(\ref{eq:1.3}) with $p+q=2$, $p+q=4$ and $p+q=6$, for which
after calculations one obtains:
\begin{eqnarray}
F_{20} & = & 1 - \frac12 M^2, \nonumber \\
F_{02} & = & 1 - \frac12  N^2, \nonumber \\
F_{11} & = & - MN, \nonumber \\
F_{40} & = & - \frac12 + M^2 - \frac14 M^4, \nonumber \\
F_{04} & = & - \frac12 + N^2 - \frac14 N^4, \nonumber \\
F_{31} & = &  2MN - M^3N, \nonumber \\
F_{13} & = &  2MN - MN^3, \nonumber \\
F_{22} & = & -1 +M^2 + N^2 - \frac32 M^2N^2, \nonumber \\
F_{60} & = & \frac13 - \frac32 M^2 + M^4 -\frac16 M^6,
\nonumber \\
F_{06} & = & \frac13 - \frac32 N^2 + N^4 -\frac16 N^6,
\nonumber \\
F_{51} & = & - 3MN + 4 M^3N - M^5N, \nonumber \\
F_{15} & = & - 3MN + 4MN^3 - MN^5,   \nonumber \\
F_{33} & = & - 6MN + 4MN^3 + 4M^3N - \frac{10}{3} M^3N^3,
\nonumber \\
F_{42} & = & 1 - \frac32 N^2 -3M^2 + M^4 + 6M^2N^2 -
\frac52 M^4N^2, \nonumber \\
F_{24} & = & 1 - \frac32 M^2 -3N^2
+ N^4 + 6M^2N^2 - \frac52 M^2N^4. \nonumber \\
\label{eq:1.4}
\end{eqnarray}
To obtain the potential generated by the whole system of the
equally oriented filaments it is necessary to sum partial
contributions (\ref{eq:1.3}) with their corresponding
coordinates $L, H$.
Below we restrict our consideration by the charge systems
constructed on the basis of the pair of parallel filaments
symmetrically situated relatively the plane $\tilde y=0$ (Fig.~\ref{fig:1}).
In particular, it includes such important cases as the
crystals with orthorhombic and tetragonal symmetry.
In the local coordinate system
presented at Fig.~\ref{fig:1} our pair has the coordinates as
$(L,H)$ and $(-L,H)$. It is obvious that the summation
of the contributions (\ref{eq:1.3}) for this pair results in
vanishing of the terms in which $M$ occurs to an odd power.
As a result, the relevant part of the electric potential
generated by the pair of the uniformly charged parallel
filaments will be:\footnote{Naturally, the same result takes
place for the pair $(L,H)$, $(L,-H)$.}
\begin{equation}
U(\tilde y, \tilde z) = U_2 (\tilde y, \tilde z) +
U_4 (\tilde y, \tilde z) + U_6 (\tilde y, \tilde z),
\label{eq:1.5}
\end{equation}
where
\begin{eqnarray}
U_2 (\tilde y, \tilde z)& = & 2\xi g_2 (L,H)(\tilde z^2 -
\tilde y^2),
\label{eq:1.6} \\
U_4 (\tilde y, \tilde z)& = & 2\xi g_4 (L,H)(\tilde y^4 -
6\tilde y^2 \tilde z^2 + \tilde z^4),
\label{eq:1.7} \\
U_6 (\tilde y, \tilde z)& = & 2\xi g_6 (L,H)\{(\tilde z^6 -
\tilde y^6) \nonumber \\
& & - 15 \tilde y^2 \tilde z^2 (\tilde z^2 - \tilde y^2)\},
\label{eq:1.8} \\
g_2 (L,H) & = & \frac{H^2 - L^2}{(H^2 + L^2)^2},
\label{eq:1.9} \\
g_4 (L,H) & = & \frac12 \frac{H^4-6H^2L^2 + L^4}{(H^2 + L^2)^4},
\label{eq:1.10} \\
g_6 (L,H) & = & \frac13
\frac{(H^6-L^6) -15H^2L^2(H^2 - L^2)}{(H^2 + L^2)^6}. \nonumber \\
& &
\label{eq:1.11}
\end{eqnarray}
The coefficient 2 is introduced into expressions (\ref{eq:1.6})%
-(\ref{eq:1.8}) to emphasize that the field in question is
assumed to be generated by the pair of the charged filaments.
Thus, the factors $g_2$, $g_4$, $g_6$ describe the contribution
from the one charged filament in given symmetric configuration.
For the system of the charged filaments with the coordinates
$(L,H)$, $(-L,H)$, $(L,-H)$, $(-L,-H)$ the coefficient 2
in (\ref{eq:1.6})-(\ref{eq:1.8}) is replaced by the factor of 4.

The results (\ref{eq:1.6})-(\ref{eq:1.11}) can be generalized
for the charge distributed in the plane
$\tilde z =H$ with the surface
density $f(L)$ which is a constant in $\tilde x$-direction
and $f(L)=f(-L)$.
Then in the formula (\ref{eq:1.6})-(\ref{eq:1.7}) the linear
charge density  $\xi$ will be replaced by $f(L)dL$.
As a result, the final expressions for the summary
contribution from this charged plane in the crystalline field
potential in the point $(\tilde x, \tilde y, \tilde z)$ look as:
\begin{equation}
U_2 (\tilde y, \tilde z) = 2G_2  \cdot (\tilde z^2 - \tilde y^2),
\label{eq:1.12}
\end{equation}
\begin{equation}
U_4 (\tilde y, \tilde z) = 2G_4 \cdot (\tilde y^4 - 6\tilde y^2
\tilde z^2 + \tilde z^4),
\label{eq:1.13}
\end{equation}
\begin{equation}
U_6 (\tilde y, \tilde z) = 2G_6 \cdot \{ (\tilde z^6 - \tilde y^6)
- 15 \tilde y^2 \tilde z^2 (\tilde z^2 - \tilde y^2) \},
\label{eq:1.14}
\end{equation}
where
\begin{equation}
G_n = G_n (H) = \int_0^\infty dL f(L) \cdot g_n (L,H).
\label{eq:1.15}
\end{equation}
In the case of two parallel identically charged planes situated
at $\tilde{z}=H$ and $\tilde{z} =-H$, the coefficient 2
in (\ref{eq:1.12})-(\ref{eq:1.14}) is replaced by 4.
The expressions (\ref{eq:1.12})-(\ref{eq:1.14}) stay formally
valid not only for two dimension
charge but for the volume one too if its distribution described
by the density $f(L,H)$, i.e. if it does not depend of
the $\tilde{x}$ -coordinate and $f(L,H)=f(-L,H)$.
For instance, it takes place when the electric field is
generated by two cords of  arbitrary cross section form.
For such a case:
\begin{equation}
G_n = \int_0^\infty dL \int dH f(L,H) \cdot g_n (L,H)
\label{eq:1.16}
\end{equation}
and again if the charge distribution is symmetric relatively
not only $\tilde{y}=0$ plane  but also $\tilde{z}=0$ plane,
the coefficient 2 in (\ref{eq:1.12})-(\ref{eq:1.14})
should be substituted by 4 and
\begin{equation}
G_n = \int_0^\infty dL \int_o^\infty dH f(L,H) g_n (L,H).
\label{eq:1.17}
\end{equation}
As in the case of the point-charge model the determination
of the perturbing Hamiltonian is the evaluation of the
appropriate electrostatic potential.
One-particle Hamiltonian for the f-electron in a crystalline
electric field:
\begin{equation}
\hat{h} = e U(\tilde y, \tilde z).
\label{eq:1.18}
\end{equation}
The calculation of the ESH parameters becomes much more
convenient when using the spherical coordinates $r, \tilde{\vartheta},
\tilde{\varphi}$ in
expression for $U(\tilde{y},\tilde{z})$ instead of the 
rectangular coordinates $x,y,z$:
\begin{equation}
\tilde x = r \sin \tilde{\vartheta} \cos \tilde{\varphi},
\quad
\tilde y = r \sin \tilde{\vartheta} \sin \tilde{\varphi},
\quad
\tilde z = r \cos \tilde{\vartheta}.
\label{eq:1.19}
\end{equation}
On proceeding of this substitution in (\ref{eq:1.5}),
(\ref{eq:1.12})-(\ref{eq:1.14}) one can obtain after
algebraic transformations the following expression for
the electrostatic potential:
\begin{equation}
U (\tilde y, \tilde z) = \sum_{n,m} \tilde{A}_{nm} r^n
Y_n^m (\tilde{\vartheta}, \tilde{\varphi}),
\label{eq:1.20}
\end{equation}
where $n=2, 4, 6$;    $m=0, \pm2, \pm4, ...\pm n$ and the
coefficients $\tilde{A}_n^m$  are equal:
\begin{eqnarray}
\tilde{A}_2^0 & = & 2G_2 \cdot 2\left ( \frac{\pi}{5} \right )^{1/2},
\nonumber \\
\tilde{A}_2^{\pm 2} & = & 2G_2 \cdot \left ( \frac{2\pi}{15}
\right )^{1/2}, \nonumber \\
\tilde{A}_4^0 & = & 2G_4 \cdot \frac23 (\pi)^{1/2}, \nonumber \\
\tilde{A}_4^{\pm 2} & = & 2G_4 \cdot \frac23 \left ( \frac{2\pi}{5}
\right )^{1/2}, \nonumber \\
\tilde{A}_4^{\pm 4} & = & 2G_4 \cdot \frac13 \left ( \frac{2\pi}{35}
\right )^{1/2}, \nonumber \\
\tilde{A}_6^0 & = & 2G_6 \cdot 2\left ( \frac{\pi}{13}
\right )^{1/2}, \nonumber \\
\tilde{A}_6^{\pm 2}&  = & 2G_6 \cdot 15\left ( \frac{2\pi}{2730}
\right )^{1/2}, \nonumber \\
\tilde{A}_6^{\pm 4} & = & 2G_6 \cdot \frac17 \left ( \frac{14\pi}{13}
\right )^{1/2}, \nonumber \\
\tilde{A}_6^{\pm 6} & = & 2G_6 \cdot \frac{1}{231} \left (
\frac{231\pi}{13} \right )^{1/2}, \nonumber \\
\label{eq:1.21}
\end{eqnarray}
It should be noted that during these manipulations we used
the notations for the spherical harmonics from~\cite{Hutchings}.

In real crystalline structures the potential
of electric field are
formed simultaneously by the ligand charges (described, for
instance, in framework of the point-charge model)
and by the extended charge structures described by the
expressions (\ref{eq:1.20})-(\ref{eq:1.21}).
The calculation of a such superposition potential and
analysis of summary system properties can be carried out
with the most convenience using the coordinate system associated
with the crystallographic symmetry axes.
Let us consider the case when the local coordinate system $\tilde{x},\tilde{y},\tilde{z}$ 
is oriented relatively the crystallographic system $x,y,z$
as at the Fig.~\ref{fig:1}.
It is the orientation that is of the most importance for
description of orthorhombic and tetragonal systems.
In the spherical coordinate system connected with the
crystallographic axes:
\begin{equation}
x = r \sin \vartheta \cos \varphi, \quad
y = r \sin \vartheta \sin \varphi, \quad
z = r \cos \vartheta.
\label{eq:1.22}
\end{equation}
Correspondingly in the local coordinate system
(see (\ref{eq:1.19}))we have:
\begin{equation}
\tilde \vartheta = \vartheta, \quad \tilde \varphi = \varphi - \Delta.
\label{eq:1.23}
\end{equation}
As a result, in (\ref{eq:1.20}) the replacement arises:
\[\tilde{A}_n^m \to A_n^m = \tilde{A}_n^m e^{- \mathrm{i} m\Delta}.\]
Then
\begin{eqnarray}
U (r,\vartheta, \varphi ) & = & \sum_{n,m} A_n^m r^n
Y_n^m(\vartheta, \varphi)  \nonumber \\
& = & \sum_{n,m} \tilde{A}_n^m \cdot e^{-\mathrm{i} m\Delta} \cdot 
r^n \cdot Y_n^m(\vartheta, \varphi),
\label{eq:1.24}
\end{eqnarray}
where the potential $U(r, \vartheta, \varphi)$ is assumed to be notated in the
crystallographic coordinates whereas the coefficients $\tilde{A}_n^m$
are calculated in the local one.
Three charge configurations with different symmetry are
shown at the Fig.~\ref{fig:2}.
Corresponding set of the $A_n^m$-coefficients are presented in
the Table~\ref{table:1}.
\begin{table}
\caption{The set of the coefficients $A_n^m$  corresponding
to the charge distribution configurations shown at
the Fig.~\ref{fig:2}.}
\label{table:1}
\begin{tabular}{lccc}
\toprule
& (a) & (b) & (c) \\
\colrule
$A_2^0$
& $4G_2 2\left (\frac{\pi}{5}\right )^{1/2}$
& $4G_2 2 \left (\frac{\pi}{5} \right )^{1/2}$
& $8G_2 2 \left (\frac{\pi}{5} \right )^{1/2}$ \\
$A_2^{\pm 2}$
& $4G_2 \left (\frac{2\pi}{15}\right )^{1/2}$
& 0
& 0 \\
$A_4^0$
& $4G_4 \frac23 (\pi)^{1/2}$
& $4G_4 \frac23 (\pi)^{1/2}$
& $8G_4 \frac23 (\pi)^{1/2}$ \\
$A_4^{\pm 2}$
& $4G_4 \frac23 \left (\frac{2\pi}{5}\right )^{1/2}$
& 0
& 0 \\
$A_4^{\pm 4}$
& $4G_4 \frac13 \left (\frac{2\pi}{35}\right )^{1/2}$
& $4G_4 \frac13 \left (\frac{2\pi}{35}\right )^{1/2}$
& $8G_4 \frac13 \left (\frac{2\pi}{35}\right )^{1/2}$ \\
$A_6^0$
& $4G_6 2 \left (\frac{\pi}{13}\right )^{1/2}$
& $4G_6 2 \left (\frac{\pi}{13}\right )^{1/2}$
& $8G_6 2 \left (\frac{\pi}{13}\right )^{1/2}$ \\
$A_6^{\pm 2}$
& $4G_6 15 \left (\frac{2\pi}{2730}\right )^{1/2}$
& 0
& 0 \\
$A_6^{\pm 4}$
& $4G_6 \frac17 \left (\frac{14\pi}{13}\right )^{1/2}$
& $4G_6 \frac17 \left (\frac{14\pi}{13}\right )^{1/2}$
& $8G_6 \frac17 \left (\frac{14\pi}{13}\right )^{1/2}$ \\
$A_6^{\pm 6}$
& $4G_6 \frac{1}{231} \left (\frac{231\pi}{13}\right )^{1/2}$
& 0
& 0 \\
\botrule
\end{tabular}
\end{table}
It should be noted that at $\Delta \not= 0, \frac{\pi}{2}$ the coefficients $A_n^m$ have
imaginary parts.

\section{Calculation of the ESH parameters}

Following the procedure proposed in~\cite{Bleaney}
for determination of the ESH parameters we shall turn
to tesseral harmonics from spherical ones in (\ref{eq:1.24}).
Accordingly~\cite{Hutchings} for even $m$ the tesseral
harmonics can be expressed via spherical harmonics as:
\begin{eqnarray}
Z_{n0} & = & Y_n^0, \nonumber \\
Z_{nm}^c & = & \frac{1}{\sqrt2} \{ Y_n^m + Y_n^{-m} \},
\nonumber \\
Z_{nm}^s & = & \frac{1}{\mathrm{i} \sqrt{2}} \{ Y_n^m -
Y_n^{-m} \},
\label{eq:2.1}
\end{eqnarray}
where $m$=2,4,6 ¨ $m \leq n$.
Accordingly:
\begin{eqnarray}
Y_n^m & = & \frac{1}{\sqrt2} \{Z_{nm}^c + \mathrm{i}
Z_{nm}^s \}, \nonumber \\
Y_n^{-m} & = & \frac{1}{\sqrt2} \{Z_{nm}^c - \mathrm{i}
Z_{nm}^s \}.
\label{eq:2.2}
\end{eqnarray}
Taking into account (\ref{eq:2.1})-(\ref{eq:2.2}), one can
obtain from (\ref{eq:1.24}):
\begin{equation}
U(r,\vartheta ,\varphi ) = \sum_n r^n \sum_{m \geq 0}
\sum_\alpha \gamma_{nm}^\alpha Z_{nm}^\alpha,
\label{eq:2.3}
\end{equation}
where $\alpha=0,c,s.$

Then for the charge system formed on the basis of the set of
parallel filaments (i.e. with the potential described in the
local coordinate system by the expressions (\ref{eq:1.21})),
we have in the crystallographic coordinate system:
\begin{eqnarray}
\gamma_{nm}^0 & = & \tilde{A}_n^0 \cdot \delta_{m0}, \nonumber \\
\gamma_{nm}^c & = & \sqrt{2} \tilde{A}_n^m \cos (m\Delta) \cdot (1-\delta_{m0}),
\nonumber \\
\gamma_{nm}^s & = & \sqrt{2} \tilde{A}_n^m \sin (m\Delta) \cdot (1-\delta_{m0}).
\label{eq:2.4}
\end{eqnarray}
There exists the universally adopted method for getting rid
of $s$-harmonics $Z_{nm}^s$  by the corresponding choice of the coordinate
system where $\gamma_{nm}^s=0$.
However, aiming to derive formulas suitable for calculation
of superposition potentials we shall not restrict extension
of our expressions.
The collection of formulas~\cite{Hutchings} expressing
the tesseral harmonics as a functions of the rectangular
coordinates is completed below with the $s$-harmonics.
Using the notations:
\begin{eqnarray}
Z_{nm}^c & = & K_{nm} \frac{f_{nm}^c}{r^n}, \nonumber \\
Z_{nm}^s & = & K_{nm} \frac{f_{nm}^s}{r^n}, \nonumber \\
Z_{nm}^0 & = & K_{n0} \frac{f_{n0}^0}{r^n},
\label{eq:2.5}
\end{eqnarray}
and the expressions (\ref{eq:1.22}), (\ref{eq:2.1}) and also
the spherical harmonics from~\cite{Hutchings}, we have obtained
the formulas for the $f_{nm}^c; f_{nm}^s$ listed in the Table~\ref{table:2}.
\begin{table*}
\caption{Components of the tesseral harmomics expressed in
Cartesian coordinates.}
\label{table:2}
\begin{tabular}{lccc}
\toprule
n,m
&$K_{nm}$
&$f_{nm}^c$
&$f_{nm}^s$ \\
\colrule
\multicolumn{1}{l} {2,0}
&\multicolumn{1}{c} {$\frac14 \left ( \frac{5}{\pi}
   \right )^{1/2}$}
&\multicolumn{2}{c} {$3z^2 - r^2$} \\
2,2
&$\frac14 \left ( \frac{15}{\pi} \right )^{1/2}$
&$x^2 - y^2$
&$2xy$ \\
\multicolumn{1}{l} {4,0}
&\multicolumn{1}{c} {$\frac{3}{16} \left ( \frac{1}{\pi}
      \right )^{1/2}$}
&\multicolumn{2}{c} {$35z^4 - 30z^2r^2 + 3r^4$} \\
4,2
&$\frac38 \left ( \frac{5}{\pi} \right )^{1/2}$
&$(7z^2 -r^2)(x^2 - y^2)$
&$(7z^2 - r^2)2xy$ \\
4,4
&$\frac{3}{16} \left ( \frac{35}{\pi} \right )^{1/2}$
&$x^4 - 6x^2y^2 + y^4$
&$4xy(x^2 - y^2)$ \\
\multicolumn{1}{l} {6,0}
&\multicolumn{1}{c} {$\frac{1}{32} \left ( \frac{13}{\pi}
      \right )^{1/2}$}
&\multicolumn{2}{c} {$231z^6 - 315z^4r^2 + 105z^2r^4 -
5r^6$} \\
6,2
&$\frac{1}{64} \left ( \frac{2730}{\pi}\right )^{1/2}$
&$\{ 16z^4 -16(x^2 + y^2)z^2 + (x^2 + y^2)^2\} (x^2 - y^2);$
&$\{ 16z^4 -16(x^2 + y^2)z^2 + (x^2 + y^2)^2\} 2xy$ \\
6,4
&$\frac{21}{32} \left ( \frac{13}{7\pi}\right )^{1/2}$
&$(11z^2 - r^2)(x^4 - 6x^2y^2 + y^4)$
&$(11z^2 - r^2)4xy(x^2 - y^2)$ \\
6,6
&$\frac{231}{64} \left ( \frac{26}{231\pi}\right )^{1/2}$
&$x^6 - 15x^4y^2 + 15x^2y^4 - y^6$
&$2xy\{4(x^2 - y^2)^2 - (x^2 + y^2)^2\}$ \\
\botrule
\end{tabular}
\end{table*}

The Hamiltonian of the f-shell in the CEF is as follows:
\begin{equation}
\hat{H}_{\text{cf}} = -|e| \sum_i U(x_i;y_i;z_i),
\label{eq:2.6}
\end{equation}
where we are normally concerned with the summation over the
f-electrons.
Thus:
\begin{equation}
\hat{H}_{\text{cf}}=-|e| \sum_n \sum_{m \geq 0} \sum_{\alpha}
\gamma_{nm}^{\alpha} \cdot K_{nm} \cdot \sum_i f_{nm}^{\alpha}(x_i;y_i;z_i),
\label{eq:2.7}
\end{equation}

Using Wigner-Eckart theorem \cite{Hutchings,Bleaney} one can
replace:
\begin{equation}
\sum_i f_{nm}^\alpha (x_i;y_i;z_i)=\Theta_n \langle r^n
\rangle \hat{O}_{nm}^\alpha,
\label{eq:2.8}
\end{equation}
Then:
\begin{equation}
\hat{H}_{\text{cf}} = \sum_n \sum_{m \geq 0} \sum_{\alpha}
B_{nm}^{\alpha} \hat{O}_{nm}^{\alpha},
\label{eq:2.9}
\end{equation}
where $\hat{O}_{nm}^{\alpha}$ - operator equivalents
introduced by Stevens \cite{Bleaney} and
\begin{equation}
B_{nm}^{\alpha}=-|e| \gamma_{nm}^{\alpha} K_{nm}\Theta_n
\langle r^n \rangle,
\label{eq:2.10}
\end{equation}

The so-called irreducible matrix elements $\Theta_n$  are
listed in \cite{Hutchings}. They are often denoted as
$\Theta_n = \alpha_J; \beta_J; \gamma_J$  for n=2, 4, 6,
respectively. The averaged value for the $n$th-power of the
f-shell radii is
denoted as $\langle r^n \rangle$  \cite{Taylor}.
Taking into account (\ref{eq:2.4}), (\ref{eq:2.10})
let us write down now the expressions for the
$B_{nm}^{\alpha}$  for the charge
system characterized by the angle of rotation $\Delta$
relatively crystallographic coordinate system as at
Fig.~\ref{fig:1}:
\begin{eqnarray}
B_{20} &=& \frac12 b_2, \nonumber \\
B_{22}^c &=& \frac12 \cos (2\Delta) b_2, \nonumber \\
B_{22}^s &=& \frac12 \sin (2\Delta) b_2, \nonumber \\
B_{40} &=& \frac18 b_4, \nonumber \\
B_{42}^c &=& \frac12 \cos (2\Delta) b_4, \nonumber \\
B_{42}^s &=& \frac12 \sin (2\Delta) b_4, \nonumber \\
B_{44}^c &=& \frac18 \cos (4\Delta) b_4, \nonumber \\
B_{44}^s &=& \frac18 \sin (4\Delta) b_4, \nonumber \\
B_{60} &=& \frac{1}{16} b_6, \nonumber \\
B_{62}^c &=& \frac{15}{32} \cos (2\Delta) b_6, \nonumber \\
B_{62}^s &=& \frac{15}{32} \sin (2\Delta) b_6, \nonumber \\
B_{64}^c &=& \frac{3}{16} \cos (4\Delta) b_6, \nonumber \\
B_{64}^s &=& \frac{3}{16} \sin (4\Delta) b_6, \nonumber \\
B_{66}^c &=& \frac{1}{32} \cos (6\Delta) b_6, \nonumber \\
B_{66}^s &=& \frac{1}{32} \sin (6\Delta) b_6,
\label{eq:2.11}
\end{eqnarray}
where:
\begin{equation}
b_n = - \tilde C |e| G_n \Theta_n \langle r^n \rangle,
\label{eq:2.12}
\end{equation}
and the coordination number $\tilde C$  is equal 2
for the charge distributed within  the $z=H$ plane with the
density $f(L)=f(-L)$, and $\tilde C =4$ when the charge is placed within
the $z= \pm H$ planes with the density
$f(L,H)=f(-L,H)=f(L,-H)=f(-L,-H)$.
Naturally, the coefficient $\tilde C$  takes the same values
for the cases described by the formulas (\ref{eq:1.15}),
(\ref{eq:1.16}) respectively.

The Stevens operator equivalents are listed for $\alpha=0;c$
in \cite{Hutchings}. The expressions for the
operators with $\alpha=s$  can be derived by means of well-known
procedure consisting of the
replacement of the coordinates $x, y, z$ in the expressions
for $f_{nm}^{\alpha}$  by the corresponding components
$J_x;J_y;J_z$  of the angular momentum operator $J$.
It is the reason why the formulas for the tesseral
harmonics in CEF-formalism are transformed back to Cartesian
coordinates. It should be noted that
it is conventional in Stevens formalism to use the operators
$\hat{J}_{\alpha}$ without $\hbar$ in the commutation rules,i.e.:
\begin{eqnarray}
\left[\hat{J}_x;\hat{J}_y \right]=\mathrm{i} \hat{J}_z,
\nonumber \\
\left[\hat{J}_y;\hat{J}_z \right]=\mathrm{i} \hat{J}_x,
\nonumber \\
\left[\hat{J}_z;\hat{J}_x \right]=\mathrm{i} \hat{J}_y,
\label{eq:2.13}
\end{eqnarray}
For the operators $\hat{J}_{\pm}$ defined as:
\begin{equation}
\hat{J}_{\pm}=\hat{J}_x \pm \mathrm{i} \hat{J}_y,
\label{eq:2.14}
\end{equation}
we have:
\begin{equation}
\hat{J}_{\pm}|J,M\rangle = \sqrt{J(J+1)-M(M \pm 1)} \cdot
|J,M \pm 1 \rangle ,
\label{eq:2.15}
\end{equation}

However, as long as the commutation rules for the operators
$\hat{x}; \hat{y}; \hat{z}$  and
$\hat{J}_x; \hat{J}_y; \hat{J}_z$  are different, the
replacement procedure demands preliminary symmetrization of
$f_{nm}^\alpha$  over all possible  permutations of
the coordinates $x, y,
z$. For $f_{22}^s, f_{44}^s, f_{66}^s$  this manipulation
is trivial. For example:
\begin{equation}
f_{22}^s=2xy=\frac{1}{2\mathrm{i}}\left\{(x+\mathrm{i} y)^2-
(x-\mathrm{i} y )^2 \right\},
\label{eq:2.16}
\end{equation}
Correspondingly:
\begin{equation}
\hat{O}_{22}^s=\frac{1}{2\mathrm{i}} \left\{\hat{J}_{+}^2 -
\hat{J}_{-}^2 \right\},
\label{eq:2.17}
\end{equation}
And:
\begin{equation}
f_{44}^s=\frac{1}{2\mathrm{i}} \left\{(x+\mathrm{i} y)^4 -
(x-\mathrm{i} y)^4 \right\},
\label{eq:2.18}
\end{equation}
\begin{equation}
\hat{O}_{44}^s=\frac{1}{2\mathrm{i}}\left\{\hat{J}_{+}^4 -
\hat{J}_{-}^4 \right\},
\label{eq:2.19}
\end{equation}
\begin{equation}
f_{66}^s=\frac{1}{2\mathrm{i}}\left\{(x+\mathrm{i} y)^6 -
(x-\mathrm{i} y)^6 \right\},
\label{eq:2.20}
\end{equation}
\begin{equation}
\hat{O}_{66}^s=\frac{1}{2\mathrm{i}}\left\{\hat{J}_{+}^6 -
\hat{J}_{-}^6 \right\},
\label{eq:2.21}
\end{equation}

However, the method that requires less tedious algebraic
manipulations  than the universally adopted
one,  may be applied to obtain the expressions for arbitrary
operators $\hat{O}_{nm}^{\alpha}$. Let us take into account
that:
\begin{equation}
f_{nm}^s=\frac{1}{K_{nm}} \frac{1}{\sqrt{2}\mathrm{i}}
\left(r^n Y_n^m-r^nY_n^{-m}\right),
\label{eq:2.22}
\end{equation}
The functions $Y_n^n;Y_n^{n-1};...Y_n^{-n}$   themselves
form the full set of the irreducible tensor operators.
Thus, accordingly Wigner-Eckart theorem the combinations
$\sum_i r_i Y_n^m$  may be directly (i.e. without
using of rectangular coordinates) substituted by the set
of the equivalent tensor operators of the same rank:
\begin{equation}
\sum_i r_i^n Y_n^m(\vartheta_i, \varphi_i)=\Theta_n \langle r^n
\rangle \hat{W}_n^m,
\label{eq:2.23}
\end{equation}

Accordingly the algorithm proposed in \cite{Watanabe},
one can choose the operator proportional to $\hat{J}_{+}^n$
as the equivalent tensor operator $\hat{W}_n^n$.
Other operators of the full set can be obtained with using the
recurrent formula:
\begin{equation}
\left[\hat{J}_{-};\hat{W}_n^m \right]=\sqrt{n(n+1)-m(m-1)}
\hat{W}_n^{m-1},
\label{eq:2.24}
\end{equation}
It is easy to see that by accepting the definition
$\hat{W}_n^n=\frac{1}{\sqrt{2}}K_{nn} \hat{J}_{+}^n$
we completely match this procedure with the formulas
(\ref{eq:2.17}), (\ref{eq:2.19}), (\ref{eq:2.21}) so that:
\begin{eqnarray}
\hat{O}_{nm}^c=\frac{1}{\sqrt{2}K_{nm}} \left(\hat{W}_n^m +
\hat{W}_n^{-m}\right), \nonumber \\
\hat{O}_{nm}^s=\frac{1}{\mathrm{i} \sqrt{2}K_{nm}}
\left(\hat{W}_n^m - \hat{W}_n^{-m}\right),
\label{eq:2.25}
\end{eqnarray}
The calculations for the $\hat{O}_{nm}^s$ yields:
%\begin{widetext}
\begin{eqnarray}
\hat{O}_{42}^s & = & \frac{1}{2\mathrm{i}} \bigl \{ \hat{J}_{+}^2
\bigl [ 7\hat{J}_z^2+14 \hat{J}_z-J(J+1)+9 \bigl ] - \nonumber \\
& & \hat{J}_{-}^2 \bigl [7\hat{J}_z^2-14\hat{J}_z-J(J+1)+9 \bigr ]
\bigr \},
\label{eq:2.26}
\end{eqnarray}
\begin{eqnarray}
\hat{O}_{62}^s & = & \frac{1}{2\mathrm{i}} \bigl \{ \hat{J}_{+}^2
\bigl [ 33\hat{J}_z^4+132 \hat{J}_z^3 - 18J(J+1) \hat{J}_z^2 +
\nonumber \\
& & 273 \hat{J}_z^2 -
36J(J+1) \hat{J}_z+282 \hat{J}_z + J^2(J+1)^2 -  \nonumber \\
& & 26J(J+1)+120 \bigr ] -
\hat{J}_{-}^2 \bigl [ 33\hat{J}_z^4-132 \hat{J}_z^3- \nonumber \\
& & 18J(J+1) \hat{J}_z^2 + 273 \hat{J}_z^2 + 36J(J+1) \hat{J}_z -
\nonumber \\
& & 282\hat{J}_z+J^2(J+1)^2-26J(J+1)+ \nonumber \\
& & 120 \bigr ] \bigr \},
\label{eq:2.27}
\end{eqnarray}
\begin{eqnarray}
\hat{O}_{64}^s & = & \frac{1}{2\mathrm{i}} \bigl \{\hat{J}_{+}^4
\bigl [11\hat{J}_z^2+44\hat{J}_z-J(J+1)+50 \bigr ] - \nonumber \\
& &\hat{J}_{-}^4
\bigl [11\hat{J}_z^2-44\hat{J}_z-J(J+1)+50 \bigr ] \bigr \},
\label{eq:2.28}
\end{eqnarray}
%\end{widetext}
Jointly with the formulas listed in \cite{Hutchings},
the expressions (\ref{eq:2.17}), (\ref{eq:2.19}),
(\ref{eq:2.21}) and (\ref{eq:2.26})-(\ref{eq:2.28}) give
the full set of Stevens operators for $n$=2, 4, 6 and even $m$.
It should be noted that our expressions (\ref{eq:2.26})
- (\ref{eq:2.28}) contain the difference of two members
corresponding to the operators $\hat{W}_n^m$ and $\hat{W}_n^{-m}$
in (\ref{eq:2.25}) respectively.
Replacing this difference by sum and the factor $1/2\mathrm{i}$
by $1/2$, one can obtain, as
it is clear from the algorithm used, the expressions for the
corresponding $\hat{O}_{nm}^c$. It is easy to make
sure after simple transformations that these coincide
with the formulas from~\cite{Hutchings}.

\section{The details of the charge distribution}

Main distinctive feature of the charge systems studied in
this paper is their extended (``filament-like'')
character, i.e. the absence of dependence for charge density
distribution in the $\tilde x$  direction. As it can
be seen from  (\ref{eq:2.11}), this fact has already estated
the  strong interrelations between the coefficients
$B_{nm}^{\alpha}$  at given $n$.
For instance, $B_{22}^c=\cos (2\Delta)
B_{20}$.

It should be stressed that these relations are
independent of the
other details of the charge
distribution. The similar relations between the
coefficients for more complex superposition systems
are demonstrated in the~\ref{table:1}.

Let us consider more closely the case when the
electric charge is distributed within the $\hat{z}=H$
plane with the surface
density $f(L)=f(-L)$, i.e. when the coefficients  $G_n$
are determined by (\ref{eq:1.15}).
 Introducing the parameter $\lambda=L/H$, one can obtain from
(\ref{eq:1.9})-(\ref{eq:1.11}), (\ref{eq:1.15}):
\begin{equation}
G_n=C_n\int_{0}^{\infty}d\lambda \, f(H\lambda) \,
\chi_n(\lambda),
\label{eq:3.1}
\end{equation}
where:
\begin{equation}
C_2=\frac{1}{H}, \qquad C_4=\frac{1}{2H^3}, \qquad
C_6=\frac{1}{3H^5},
\label{eq:3.2}
\end{equation}
\begin{eqnarray}
\chi_2(\lambda) & = & \frac{1-\lambda^2}{(1+\lambda^2)^2},
\nonumber \\
\chi_4(\lambda) & = & \frac{1 - 6\lambda^2 + \lambda^4}{(1 +
\lambda^2)^4},
\nonumber \\
\chi_6(\lambda)& = & \frac{1 - \lambda^6 - 15\lambda^2(1 -
\lambda^2)}{(1 + \lambda^2)^6},
\label{eq:3.3}
\end{eqnarray}

The formulas (\ref{eq:3.1})-(\ref{eq:3.3})) describe the
contributions of the ``partial filaments'' with the coordinates
$L=H\lambda$  into the potentials of the 2-nd, 4-th and 6-th
orders respectively. The functions $\chi_n(\lambda)$ play
the role of the weighting factors.
It is clear from Fig.\ref{fig:3}
that these contributions oscillate with $\lambda$  not
synchronously. Moreover, the rates of their damping are also
sufficiently different.

At last for a certain practical calculations it is
rather convenient to define the parameter $t$ as follows:
\begin{equation}
L=H \tan (\frac{t}{2}),
\label{eq:3.4}
\end{equation}
Then:
\begin{equation}
G_n=C_n \int_0^{\pi}dt \, f(H \tan(\frac{t}{2})) \cdot \Omega_n(t),
\label{eq:3.5}
\end{equation}
\begin{equation}
\Omega_2(t)=\cos (t),
\label{eq:3.6}
\end{equation}
\begin{equation}
\Omega_4(t)=\frac14 \left \{\cos (3t) + 2\cos (2t) +
cos (t) \right \},
\label{eq:3.7}
\end{equation}
\begin{eqnarray}
\Omega_6(t) & = & \frac{1}{16} \{\cos (5t) + 4\cos (4t) +
6\cos (3t) +  \nonumber \\
& & 4\cos(2t) + \cos (t) \},
\label{eq:3.8}
\end{eqnarray}
The plots for the functions $\Omega_n(t)$ are shown at
Fig.~\ref{fig:4}.

\section{The Model of the charged stripes}

Let us consider the case when the electric charge is
distributed
along  the plane $\tilde z=H$   within
the system of the parallel stripes uniformly charged with the
surface density $F_0$. As usual, it is
assumed that $f(L)=f(-L)$.
The contribution into $G_n$ from the stripe situated
in the range
$L_N \, \leq L \, \leq (L_N+W_N)$  can be calculated using
(\ref{eq:3.1}) or (\ref{eq:3.5}). As a result one obtains:
\begin{equation}
G_n^{(N)}=F_0 C_n \left \{\nu_n(\beta_N)-\nu_n(\alpha_N) \right \},
\label{eq:4.1}
\end{equation}
\begin{eqnarray}
\nu_2(\lambda) & = & \frac{\lambda}{1+\lambda^2}, \nonumber \\
\nu_4(\lambda)& = & \frac{\lambda (3 - \lambda^2)}{3(1 +
\lambda^2)^3}, \nonumber \\
\nu_6(\lambda) & = & \frac{\lambda (5 - 10 \lambda^2 +
\lambda^4)}{5(1 + \lambda^2)^5},
\label{eq:4.2}
\end{eqnarray}
where the variables
\begin{equation}
\alpha_N=\frac{L_N}{H}, \qquad \beta_N=\frac{L_N+W_N}{H},
\label{eq:4.3}
\end{equation}
are the values of the parameter $\lambda$
corresponding to the stripe left and right edges.
The summary contribution from the total system of
the charged stripes can be calculated accordingly:
\begin{equation}
G_n=\sum_N G_n^{(N)},
\label{eq:4.4}
\end{equation}
At last it is useful to write down the final formula for
the coefficients $b_n$ taking into account the
contribution from the N-th stripe.
From (\ref{eq:2.12}) and (\ref{eq:4.1}) it follows:
\begin{equation}
b_n^{(N)}=-\tilde C e^2 \sigma \langle r^n \rangle \Theta_n C_n
\left \{\nu_n(\beta_N)-\nu_n(\alpha_N) \right \},
\label{eq:4.5}
\end{equation}

The symbol $\sigma$ here means the surface concentration
of the carriers generating the stripe charge, so that:
\begin{equation}
F_0=\sigma |e|,
\label{eq:4.6}
\end{equation}
where $e$ is the elementary charge and the variable $\sigma$
is positive for the holes and negative for the
electrons. It is convenient during the practical calculations
to replace in (\ref{eq:4.6}) the value $e^2$ by
$e^2=14400$ meV $\cdot$ 1 \AA, simultaneously expressing all the
lengths in Angstroms.
The coefficient $\tilde C$ in (\ref{eq:4.5}) is of the same
sense as in (\ref{eq:2.12}).
Carrying out summation of the expression (\ref{eq:4.5}) over
N with corresponding values of the angle $\Delta$  in
(\ref{eq:2.11}), one can obtain the formulas for the ESH
parameters in the cases of rather complex superposition
systems.

\section{Periodical two dimensional lattices of the charged
stripes}

Now we consider the case when the surface charge density
$f(L)$ in (\ref{eq:1.15}) is a periodic
function with the period $T$.
\begin{equation}
f(L+T)=f(L).
\label{eq:5.1}
\end{equation}
Introducing the surface density $\sigma(L)$ of the number
of the carriers:
\begin{equation}
f(L)=|e| \sigma(L).
\label{eq:5.2}
\end{equation}
Expanding $\sigma(L)$ to Fourier series  results in:
\begin{equation}
\sigma (L)=\sigma_0+\sum_{k=1}^{\infty} \sigma_k
\cos (\frac{2\pi k}{T}L).
\label{eq:5.3}
\end{equation}

The term $\sigma_0$  in (\ref{eq:5.3}) corresponds to the
uniformly charged plane. This part of the charge density
generates only the uniform electric field with the constant
gradient and naturally with the zero
contributions of the 2-nd, 4-th and 6-th orders into the
potential. It is easy to see from (\ref{eq:3.5})-(\ref{eq:3.7})
that our results satisfy to this condition.
Then combining (\ref{eq:3.1}) and (\ref{eq:5.3}):
\begin{equation}
G_n=|e| C_n \sum_{k=1}^{\infty} E_{nk} \sigma_k,
\label{eq:5.4}
\end{equation}
where the coefficients
\begin{equation}
E_{nk}=\int_0^{\infty}d\lambda \, \chi_n(\lambda)
\cos (2 \pi k \frac{H}{T} \lambda)
\label{eq:5.5}
\end{equation}
depend only on the relation $H/T$ and are not connected
with the other details of the charge
distribution described by the coefficients $\sigma_k$.
The formulas (\ref{eq:4.5}) stay to be valid for the system
in question but now they look as:
\begin{equation}
b_n=-\tilde C e^2 C_n \Theta_n \langle r^n \rangle
\sum_{k=1}^{\infty} E_{nk} \sigma_k,
\label{eq:5.6}
\end{equation}
As it is clear from  (\ref{eq:4.5}), the full set of the
coefficients $B_{nm}^{\alpha}$  within the framework of the
model developed is completely defined by the values of
the three parameters $b_2, \, b_4, \, b_6$, which can be
obtained, for example, by fitting of the experimental and
calculated neutron spectra. Thus, the
experimental information can be a tool for verification both
of the microscopic and of the
phenomenological models of the charge distribution with
three arbitrary parameters.
It should be noted that the expressions (\ref{eq:5.4}) -
(\ref{eq:5.6}) describe the contributions from one or two
charged planes dependingly of the coefficient $\tilde C$
($\tilde C=2 $  or $4$, respectively).
But they can be easily
generalized for more complicate configurations.
For instance, in the case of a stack of identically
charged planes, in (\ref{eq:5.6}) one should estate
$\tilde C=4$  and  substitute the coefficients $E_{nk}$  by
the $\tilde E_{nk}$:
\begin{equation}
\tilde E_{nk}=\sum_H E_{nk},
\label{eq:5.7}
\end{equation}
where the summation of (\ref{eq:5.5}) is carried out over the
plane coordinates $H>0$.

Now a few words about the periodicity of the charge
distribution. Well-defined character of
the experimental neutron spectra gives us the evidence of the
fact that the crystal lattice and charge
distribution periods are commensurate.
In other case the spectrum would be smoothed.
Moreover, the coefficient interrelating these periods should
be not large due to the same reasons.
There are three examples shown at the Fig.\ref{fig:5}.
\begin{itemize}
\item The period of the charge distribution is equal to the
lattice one. In such a case we have identical
positions for every Rare-Earth ion and only one set of the
crystalline electric field parameters.
\item If $T=2a$, the situation is different.
We have two types of the REI positions with their own crystal
field. If the condition $f(L)=f(-L)$ is satisfied then both
positions are symmetric.
\item If $T=3a$ there exist three types of positions.
But if we deal with the systems with at least
orthorhombic symmetry then the charge distribution in the
elementary cell should be rather symmetric too.
In other case the distortions seem to be
inevitable. The condition $f(L)=f(-L)$
guaranties, at least, the symmetry of distribution for one
type of positions. Then immediately we get
rather interesting thing: there are two different positions
("1" and "2") for the REI in such a system. But the "1"
position is asymmetric. As a consequence there must be the
$B_{nm}$ coefficients with both even and odd $m$.
\end{itemize}

As a rather general  example, let us consider the model of
two parabolas (Fig.\ref{fig:6}) when the
period $T=a$ and the function $\sigma (L)$ in the range of
 $(0, \, \frac{a}{2})$ is assumed to be equal to:
\begin{equation}
\sigma (L)=\sigma_0 + \kappa L^2 + \Theta (L-L_0)
\frac{\mu}{\delta} \left \{\delta^2 - (L-\frac{a}{2})^2
\right \},
\label{eq:5.8}
\end{equation}
where $L_0=\frac{a}{2}-\delta$;  $\sigma_0$ is the constant
that does not affect  $b_2, \, b_4, \, b_6$  and should be
obtained by
means of  a certain additional speculations
\[
\Theta (L -L_0) = \left \{
\begin{array}{ll}
1, \quad L \geq L_0, \\
0, \quad L < L_0.
\end{array}
\right.
\]
Thus, accordingly this model the function $\sigma(L)$ in
the range of $(0,\frac{a}{2})$  is a superposition of the two
parabolas with the extremums in the points $L=0$ and
$L=a/2$ respectively.
The coefficients $\kappa$ and $\mu$  can
be of arbitrary signs. Also taking into account the
possibility of adding of an arbitrary constant to
(\ref{eq:5.8}), one can see that this formula describes
rather wide set of the charge distributions.
After  calculation of the Fourier coefficients for the
function (\ref{eq:5.8}) we obtain:
\begin{eqnarray}
\sigma_k & = & (-1)^k \frac{a}{\pi^2 k^2} \left \{a \kappa +
2 \mu \left [\frac{a}{2\pi k \delta} \sin (\frac{2\pi k
\delta}{a}) - \right . \right . \nonumber \\
& & \left . \left . \cos (\frac{2\pi k \delta}{a}) \right ]
\right \},
\label{eq:5.9}
\end{eqnarray}
As a result, the system of the three equations
(at $n=2, \, 4, \, 6$) interrelating the variables
$\kappa, \mu, \delta$  may be
derived from (\ref{eq:5.6}), (\ref{eq:5.9}):
\begin{equation}
P_n=\kappa Q_n + \mu S_n(\delta),
\label{eq:5.10}
\end{equation}
where:
\begin{equation}
P_n=-\frac{b_n \pi^2}{\tilde C e^2 \Theta_n \langle r^n
\rangle C_n a},
\label{eq:5.11}
\end{equation}
\begin{equation}
Q_n=a \sum_{k=1}^{\infty} E_{nk} \frac{(-1)^k}{k^2},
\label{eq:5.12}
\end{equation}
\begin{equation}
S_n(\delta)=2 \sum_{k=1}^{\infty} E_{nk} \frac{(-1)^k}{k^2}
\left [\frac{a}{2\pi k \delta} \sin (\frac{2\pi k \delta}{a}) -
\cos (\frac{2\pi k \delta}{a}) \right ].
\label{eq:5.13}
\end{equation}
The equations (\ref{eq:5.10}) are linear relatively the
variables $\kappa, \mu$. The solution of this system easily
can be
reduced to determination of $\delta$ from the following
transcendent equation:
\begin{eqnarray}
& & (Q_2 P_4 - Q_4 P_2) S_6(\delta) + (Q_6 P_2 - Q_2 P_6)
S_4(\delta)  +  \nonumber \\
& & (Q_4 P_6 - Q_6 P_4) S_2(\delta) = 0.
\label{eq:5.14}
\end{eqnarray}
After solving of (\ref{eq:5.14}), other parameters will
be determined immediately:
\begin{eqnarray}
\mu=\frac{Q_2 P_4 - Q_4 P_2}{Q_2 S_4(\delta)-
Q_4 S_2(\delta)}, \nonumber \\
\kappa=\frac{P_2 S_4(\delta) - P_4 S_2(\delta)}{Q_2 S_4(\delta)
- Q_4 S_2(\delta)},
\label{eq:5.15}
\end{eqnarray}

Sometimes from certain additional speculations (for example,
concerning the concentration of
dopants in the crystalline lattice) one can independently
indicate the concentration of the carriers
injected into the plane $z=H$. Let us denote the total number
of these carriers per the element $a \times d$  of
the charged plane shown at the Fig.\ref{fig:6} as $N_0$.
In such a case the expressions (\ref{eq:5.14}) -
(\ref{eq:5.15}) can be
completed by the formulas for determination of the
$\sigma_0$ term from (\ref{eq:5.8}). Really:
\begin{equation}
N_0 = 2d \int_{0}^{\frac{a}{2}} dL \sigma (L),
\label{eq:5.16}
\end{equation}
Substituting (\ref{eq:5.8}) into (\ref{eq:5.16}) yields:
\begin{equation}
\sigma_0 = \frac{N_0}{ad} - \frac{1}{12}a^2 \kappa -
\frac{8}{3a} \mu \delta^2,
\label{eq:5.17}
\end{equation}
Sometimes the calculation of $b_n$  is convenient not making
use of Fourier expansions. For example,
direct summation of the expressions (\ref{eq:4.5}) over $N$
is possible for the model of the charged stripes.
Accordingly (\ref{eq:2.12}):
\begin{equation}
b_n = - \tilde C \cdot |e| \cdot \Theta_n \cdot \langle r^n \rangle
\sum_{N=1}^{\infty} G_n^{(N)},
\label{eq:5.18}
\end{equation}
For the periodical lattice of the stripes charged with
the surface density $F_0=\sigma |e|$  we have:
$f(H\lambda)=F_0$  at $\lambda \in (\alpha_N; \beta_N)$
and $f(H\lambda)=0$  at $\lambda \not\subset (\alpha_N;
\beta_N)$,
where
\begin{eqnarray}
\alpha_N = (Na - \frac{a+W}{2})/H \nonumber \\
\beta_N = (Na - \frac{a-W}{2})/H,
\label{eq:5.19}
\end{eqnarray}
$W$ is the width of a stripe.

It is possible to accelerate the convergence of the series
in (\ref{eq:5.18}) using the fact that one can add an
arbitrary constant to the charge density in the expression
(\ref{eq:3.1}). For example:
\begin{equation}
G_n=C_n \int_0^{\infty} d\lambda \cdot \left [f(H \lambda)-
\frac{F_0 W}{a} \right ] \cdot \chi_n(\lambda),
\label{eq:5.20}
\end{equation}
where we have extracted from the charge density its average
value.
In practice, the summation in (\ref{eq:5.18}) is carried out
up to a certain maximal stripe number $N=M$.
Representing (\ref{eq:5.20}) as:
\begin{eqnarray}
G_n &=& C_n \cdot \int_0^{\beta_M} d\lambda \cdot \left [
f(H\lambda)
- \frac{F_0 W}{a} \right ] \cdot \chi_n(\lambda) + \nonumber \\
&+& C_n \cdot \int_{\beta_M}^{\infty} d\lambda \cdot \left [
f(H\lambda)
- \frac{F_0 W}{a} \right ] \cdot \chi_n(\lambda),
\label{eq:5.21}
\end{eqnarray}
we then take into account that at large $\beta_M$ , as it
can be seen from the Fig.\ref{fig:3}, the second integral
contains the product of a slow varying function and the
function
oscillating around the zero average
value. It gives us the possibility to neglect the second
integral (from the physical point of view it
corresponds to the mutual compensation of the two neighboring
stripes
carrying  the charges of the
equal value but of the opposite signs).
Then calculating the first integral we obtain:
\begin{equation}
G_n \approx C_n F_0 \left [\sum_{N=1}^M \{\nu_n(\beta_N)-
\nu_n(\alpha_N) \}
- \frac{W}{a} \nu_n(\beta_M) \right ].
 \label{eq:5.22}
\end{equation}

Correspondingly:
\begin{eqnarray}
b_n & = & -\tilde C e^2 \sigma \Theta_n \langle r^n \rangle C_n
\left [\sum_{N=1}^M \{\nu_n(\beta_N)- \right .\nonumber \\
& & \left . \nu_n(\alpha_N) \} -
\frac{W}{a}
\nu_n(\beta_M) \right ].
\label{eq:5.23}
\end{eqnarray}

\section{Point Charge Model Calculation of the ESH Parameters
for the Complex Charge Configurations}

Below we list the formulas adopted for numerical calculation
of the coefficients $B_{nm}^{\alpha}$
generated by the clouds of charged points.
Then we use the results obtained for evaluation of an
applicability of the charged filaments model to description
of real systems.
In the point-charge model the coefficients
$\gamma_{nm}^{\alpha}$  are expressed via the spherical
coordinates
$R_j, \vartheta_j, \varphi_j$ of the charges generating the
crystalline electric field as \cite{Hutchings}:
\begin{equation}
\gamma_{nm}^{\alpha}=\sum_j \frac{4\pi q_j}{2n+1}
\frac{Z_{nm}^{\alpha}(\vartheta_j;\varphi_j)}{R_j^{n+1}}.
\label{eq:6.1}
\end{equation}

The necessary expressions for the tesseral harmonics are
listed in Table \ref{table:2}.  In this section we use for
the Cartesian coordinates of the charges the traditional
capital letters $X_j, Y_j, Z_j$  instead of $D_j, L_j, H_j$
that are reserved only for filaments. As a result, one can
obtain from (\ref{eq:2.10}) and Table \ref{table:2} the
following list of the formulas:
\begin{widetext}

\[ B_{20}  =-|e| \frac14 \Theta_2 \langle r^2 \rangle \sum_j q_j
\frac{3Z_j^2-R_j^2}{R_j^5}, \]
\[
B_{22}^c=-|e| \frac34 \Theta_2 \langle r^2 \rangle \sum_j q_j
\frac{X_j^2-Y_j^2}{R_j^5}, \]
\[B_{22}^s=-|e| \frac34 \Theta_2 \langle r^2 \rangle \sum_j q_j
\frac{2X_jY_j}{R_j^5}, \]
\[B_{40}  =-|e| \frac{1}{64} \Theta_4 \langle r^4 \rangle
\sum_j q_j \frac{35Z_j^4-30Z_j^2R_j^2+3R_j^4}{R_j^9},
\]
\[B_{42}^c=-|e| \frac{5}{16} \Theta_4 \langle r^4 \rangle
\sum_j q_j \frac{(7Z_j^2-R_j^2)(X_j^2-Y_j^2)}{R_j^9},\]
\[B_{42}^s=-|e| \frac{5}{16} \Theta_4 \langle r^4 \rangle
\sum_j q_j \frac{(7Z_j^2-R_j^2)2X_jY_j}{R_j^9}, \]
\[B_{44}^c=-|e| \frac{35}{64} \Theta_4 \langle r^4 \rangle
\sum_j q_j \frac{X_j^4-6X_j^2Y_j^2+Y_j^4}{R_j^9}, \]
\[B_{44}^s=-|e| \frac{35}{64} \Theta_4 \langle r^4 \rangle
\sum_j q_j \frac{4X_jY_j(X_j^2-Y_j^2)}{R_j^9}, \]
\[B_{60}  =-|e| \frac{1}{256} \Theta_6 \langle r^6 \rangle
\sum_j q_j \frac{231Z_j^6-315Z_j^4R_j^2+105Z_j^2R_j^4-
5R_j^6}{R_j^{13}}, \]
\[B_{62}^c=-|e| \frac{105}{512} \Theta_6 \langle r^6 \rangle
\sum_j q_j \frac{\left [16Z_j^4-16(X_j^2+Y_j^2)Z_j^2+(X_j^2+
Y_j^2)^2 \right ](X_j^2-Y_j^2)}{R_j^{13}}, \]
\[B_{62}^s=-|e| \frac{105}{512} \Theta_6 \langle r^6 \rangle
\sum_j q_j \frac{\left [16Z_j^4-16(X_j^2+Y_j^2)Z_j^2+(X_j^2+
Y_j^2)^2 \right ]2X_jY_j}{R_j^{13}}, \]
\[B_{64}^c=-|e| \frac{63}{256} \Theta_6 \langle r^6 \rangle
\sum_j q_j \frac{(11Z_j^2-R_j^2)(X_j^4-6X_j^2Y_j^2+Y_j^4)}
{R_j^{13}}, \]
\[B_{64}^s=-|e| \frac{63}{256} \Theta_6 \langle r^6 \rangle
\sum_j q_j \frac{(11Z_j^2-R_j^2)4X_jY_j(X_j^2-Y_j^2)}{R_j^{13}},
\]
\[B_{66}^c=-|e| \frac{231}{512} \Theta_6 \langle r^6 \rangle
\sum_j q_j \frac{X_j^6-15X_j^4Y_j^2+15X_j^2Y_j^4-
Y_j^6}{R_j^{13}}, \]
\begin{equation}B_{66}^s=-|e| \frac{231}{512} \Theta_6
\langle r^6
\rangle \sum_j q_j \frac{2X_jY_j \left [4(X_j^2-Y_j^2)^2-(X_j^2
+Y_j^2)^2 \right ] }{R_j^{13}}.
\label{eq:6.2}
\end{equation}
\end{widetext}

The expressions (\ref{eq:6.2}) are convenient for numerical
calculations in the models representing the
electric charge distributed in the lattice as a net of the
charged points.
Let us use this formulas for evaluation of the inaccuracy
induced by the replacement of a finite
filament by an infinite one.
As it can be seen from (\ref{eq:6.2}), the coefficients
$B_{nm}^{\alpha}$  decrease with a distance
as:
\begin{equation}
B_{2m}^{\alpha} \sim R_j^{-3}, \quad  B_{4m}^{\alpha}
\sim R_j^{-5}, \quad  B_{6m}^{\alpha} \sim R_j^{-7}.
\label{eq:6.3}
\end{equation}

Thus, it is obvious that the replacement in question
contributes the maximal error at $n=2$. So, it is
reasonable for our evaluation to calculate the coefficients
$B_{20}$ and $B_{22}^c$  for the finite charged filament
(exactly speaking, belonging to the configuration of the
filaments in the sense of the formulas
(\ref{eq:2.12}) oriented along the x-axis in the range
$(-D; +D)$ and characterized by the coordinates $Y=L$,
$Z=H$.  Such a configuration automatically gives $B_{nm}^s=0$.
Replacing $q_j \rightarrow \xi dx$  and transforming
(\ref{eq:6.2}) from the sum to the integral yield:
\begin{eqnarray}
B_{20} & = & -|e| \cdot \xi \cdot \Theta_2 \cdot \langle r^2 \rangle \cdot \frac12 \cdot
\frac{\sin (\varphi_0)}{(H^2+L^2)^2} \left \{(H^2-L^2)+
\right . \nonumber \\
& & \left . H^2 \cos^2 (\varphi_0) \right \},
\label{eq:6.4}
\end{eqnarray}
\begin{eqnarray}
B_{22}^c & = & -|e| \cdot \xi \cdot \Theta_2 \cdot \langle r^2 \rangle \cdot \frac14 \cdot
\frac23 \cdot \frac{\sin (\varphi_0)}{(H^2+L^2)}  \left \{\sin^2 (\varphi_0)
\right . \nonumber \\
& & \left . - \frac{L^2}{L^2+H^2} (3- \sin^2 (\varphi_0)) \right \},
\label{eq:6.5}
\end{eqnarray}
\begin{equation}
\varphi_0=\arctan (\frac{D}{\sqrt{H^2+L^2}}).
\label{eq:6.6}
\end{equation}

For the infinite filament $\varphi_0=\frac{\pi}{2}$.
Assuming $\sqrt{H^2+L^2} \ll D$, let us expand the
expressions (\ref{eq:6.4})-(\ref{eq:6.5}) over this parameter.
As a
result, the following convenient formulas for evaluation of
relative inaccuracy appear:
\begin{equation}
B_{20}=B_{20}(\infty) \cdot \left \{1-\frac12 \cdot
\frac{H^2+L^2}{L^2-H^2} \cdot \frac{H^2+L^2}{D^2} \right \},
\label{eq:6.7}
\end{equation}
\begin{equation}
B_{22}^c=B_{22}^c(\infty) \cdot \left \{1+\frac32 \cdot
\frac{H^2+L^2}{L^2-H^2} \cdot \frac{H^2+L^2}{D^2} \right \}.
\label{eq:6.8}
\end{equation}
where $B_{20}(\infty),B_{22}^c(\infty)$ - are the corresponding
expressions for the infinite filament. Naturally the
formulas (\ref{eq:6.7})-(\ref{eq:6.8}) are not applicable for
the case $L=H$ when $B_{20}(\infty),B_{22}^c(\infty)$ are equal
to zero. In
the latest case only absolute evaluations on the basis of the
expressions (\ref{eq:6.4})-(\ref{eq:6.5}) are possible.
At $L=0$ the value of the ratio $D/H=10$ is follows from
(\ref{eq:6.7})-(\ref{eq:6.8}) to be enough to provide the
accuracy of 1\% for the model of the infinite filament.
To achieve the same accuracy at $L \sim 5H$ one
needs of the ratio $D/H \sim 50$.
However, when the lattice of the stripes is described,
such an increase
concerns only the relative error of the far stripes input
whereas the main contribution to the potential
is generated by the nearby ones.

\section{Conclusion}

The new model is developed to strengthen the possibilities
for analysis of experimental data
obtained by inelastic neutron scattering.
It is particularly valuable for studying the charge
distribution in the layered perovskite-like compounds doped
with the rare-earth ions being used as a
local probe.
In contrast to the widely adopted point-charge model, the
new approach directly takes into
account the extended character of the charge distribution
in one direction.
The model provides clear analytical expressions for the
crystalline electric field parameters,
which may be used efficiently in further evaluations.
As a result, the strong interrelations between
these parameters were found to originate from the specific
charge distribution topology. The similar
relations may easily be established for more complicated,
superpositional systems.
The model enabled us to develop methods of the charge profile
reconstruction from the experimental
data.

\begin{acknowledgments}
Financial support by the Russian Foundation for Basic Research
(project No 00-02-17370), INTAS project No 99-00256,
and the State Contract No 107-19(00)-P is gratefully acknowledged.
\end{acknowledgments}

%\bibliography{cfe}
%\end{document}

\begin{figure}
\caption{The local $\tilde x; \tilde y; \tilde z$ and
crystallographic $x; y; z$ coordinate systems for description
of the field generated in the point $\mathbf{r}$ by the charged
filaments $AB$ and $A\prime B\prime$.}
\label{fig:1}
\end{figure}

\begin{figure}
\caption{ Superpositional charged systems characterized
by three parameters: (a) - symmetry $mmm (D_{2h})$;
(b) - symmetry $\overline{4}2m (D_{2d})$;
(c) - symmetry $4/mmm (D_{4h}$.}
\label{fig:2}
\end{figure}

\begin{figure}
\caption{Weighting functions $\chi_n(\lambda)$ for the
contributions to the crystalline field potential of the
2, 4 and 6 orders, respectively.}
\label{fig:3}
\end{figure}

\begin{figure}
\caption{Weighting functions $\Omega_n(t)$ for the contributions to the
crystalline field potential of the 2, 4 and 6 orders, respectively.}
\label{fig:4}
\end{figure}

\begin{figure}
\caption{Schematic charge distribution profiles with
different periodicity.}
\label{fig:5}
\end{figure}

\begin{figure}
\caption{The model of two parabolas. The summary carrier
concentration $\sigma (L)$ consists of the two prabolic
contributions $\sigma_1(L)$ and $\sigma_2(L)$ and may
be added by the constant $\sigma_0$.}
\label{fig:6}
\end{figure}

\end{document}